\documentclass[prl,twocolumn,superscriptaddress]{revtex4-2}
\usepackage[utf8]{inputenc}
\usepackage{amsmath}
\usepackage{amssymb}
\usepackage{bm}
\usepackage{graphicx}
\usepackage{hyperref}
\usepackage{braket}
\usepackage{mathtools}
\usepackage{xspace}
\usepackage{xcolor}
\definecolor{blue}{rgb}{0,0,0}
\definecolor{red}{rgb}{0,0,0}
\usepackage{physics}
\usepackage{bbm}

\begin{document}
\title{Genuine quantum scars in Floquet chaotic many-body systems}
\author{Harald Schmid}
\affiliation{\mbox{Physics Department, TUM School of Natural Sciences, Technical University of Munich, 85748 Garching, Germany}}
\affiliation{Munich Center for Quantum Science and Technology (MCQST), Schellingstr. 4, 80799 M{\"u}nchen, Germany}
\author{Andrea Pizzi}
\affiliation{Cavendish Laboratory, University of Cambridge, Cambridge CB3 0HE, United Kingdom}
\author{Johannes Knolle}
\affiliation{\mbox{Physics Department, TUM School of Natural Sciences, Technical University of Munich, 85748 Garching, Germany}}
\affiliation{Munich Center for Quantum Science and Technology (MCQST), Schellingstr. 4, 80799 M{\"u}nchen, Germany}
\begin{abstract}
Unstable periodic orbits act as organizing structures for classical chaotic systems and underpin quantum scarring. Long known in single-particle systems, genuine quantum scars based on unstable periodic orbits have been recently extended to isolated many‑body systems for time-independent Hamiltonians. Their fate under periodic driving, however, remains largely uncharted, challenged by the expectation that these systems should in general heat to infinite temperature. Here, we investigate how genuine scarring competes with the drive in a Floquet many‑body system. Using chaotic spin chains, we demonstrate that Floquet states remain scarred in the high-frequency limit. Beyond this static correspondence, we uncover additional, driving‑induced Floquet scars with no static analog. We construct a rich dynamical stability diagram with intermediate-frequency regimes of enhanced and quenched scarring, which we understand with a classical analysis of the Lyapunov exponent. Our results position Floquet systems as a natural platform for tuning the scarring behavior of quantum many-body systems.
\end{abstract}

\date{\today}
\maketitle

{\em Introduction.---} In chaotic quantum systems, scarred eigenstates feature an anomalously large probability density along certain isolated unstable periodic orbits (UPOs) of the underlying classical dynamics \cite{heller_bound-state_1984,berry_quantum_1989,kaplan_scars_1999}. A striking dynamical effect of scarring is an enhanced long-time return probability to the orbit, which is due to interference and surprising given the orbit’s classical instability. Scarring has a long history in semiclassical single-particle systems~\cite{heller_bound-state_1984}, where it has also been shown to occur in the presence of a periodic drive \cite{virovlyansky_manifestation_2005,timberlake_localization_2005}, such as in the kicked top \cite{kus_quantum_1991,dariano_classical_1992,mondal_dynamical_2021}.

\begin{figure*}[t!]
    \centering
\includegraphics[width=\linewidth]{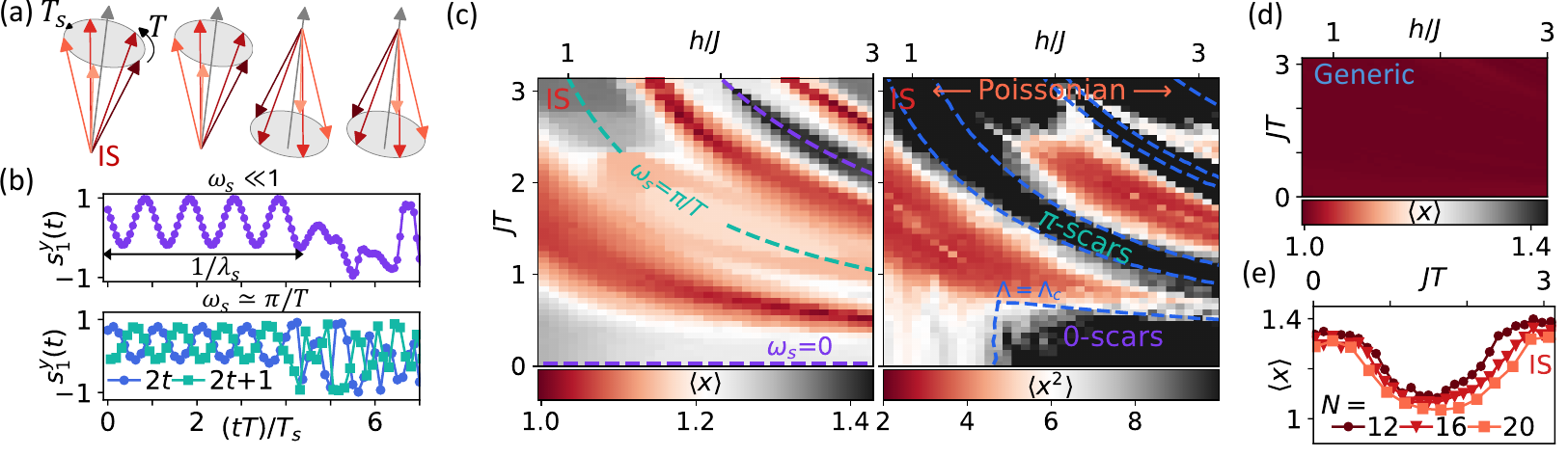}
    \caption{\textbf{Genuine scarring in a Floquet spin chain.} (a) IS states precess with period $T_s=\frac{2\pi}{\omega_s}$ which generally differs from the driving period $T$. 
    (b) Classical stroboscopic Floquet dynamics: the Lyapunov exponent $\lambda_s$ governs the instability of weakly perturbed IS states. States near  $\omega_s\simeq \pi/T $ mimic those near {\color{blue}$\omega_s/\Omega \simeq 0$} {\color{blue} over even/odd (blue/turquoise) timesteps}. 
    {\color{blue}(c)} Stability diagram of quantum Floquet scars. Mean overlap $x=|\langle \psi|E_n\rangle|^2$ for IS states exceeds the RMT value  $\langle x_{\mathrm{PT}} \rangle=1$ in the scarred regions. $0$-scars appear near $\omega_s= 0$ (dashed purple), $\pi$-scars near $\omega_s= \pi/T$ (dashed turquoise).  For the mean-squared overlaps, RMT predicts  $\langle x_{\mathrm{PT}}^2 \rangle=2$.  \textcolor{blue}{Dashed blue line: classical scarring indicator $\Lambda=\Lambda_c$.} 
    (d) Mean overlap for GR states remains close to RMT across parameter space.
    (e) Finite-size scaling of the mean overlap for IS scars (fixed $h=0.9J$).
    Parameters: $\mathcal{N}=10^{3}$ state samples, (c,d) $N=16$.}
    \label{fig:phase_diag}
\end{figure*}

The concept of scarring has more recently been extended to many-body systems. On the one hand, most work in this area has focused on selected models in which some eigenstates are strongly atypical, namely, non-thermal and area-law entangled~\cite{bernien_probing_2017,turner_weak_2018,ho_periodic_2019,serbyn_quantum_2021,moudgalya_quantum_2022,bluvstein_controlling_2021,maskara_discrete_2021,hudomal_driving_2022,mizuta_exact_2020,Russomanno2022,ScarFinder2025,Petrova2025}. These states have often been associated with underlying \textit{stable} classical periodic orbits \cite{michailidis_slow_2020,turner_correspondence_2021,lerose_theory_2025,omiya_quantum_2025,muller_semiclassical_2024,kerschbaumer_quantum_2025}. On the other hand, scars linked to \textit{unstable} periodic orbits, in closer analogy with the single-particle prescription and hence dubbed ``genuine'', was recently established for generic (possibly volume-law) eigenstates of time-independent many-body systems~\cite{Hummel2023,evrard_quantum_2024,pizzi_genuine_2025}. In the case of spins, a general recipe for genuine scarring consists in finding special \textit{interaction-suppressing} (IS) spin configurations that cancel the classical exchange fields and lead to classical precession UPOs [Fig.\ \ref{fig:phase_diag}(a)]. In some cases, IS configurations can even generate an extensive manifold of such UPOs, akin to groundstates in frustrated magnetism, but in the middle of the many-body spectrum~\cite{pizzi_unstable_2025}.

The syndrome of genuine scarring consists of atypical eigenstates' morphology, amplitude statistics, and long-time return probabilities, which can be viewed as curbing chaos in these systems. For this reason, observing genuine scarring in a many-body system under a periodic drive would be even more remarkable, as the drive tends to heat the system and typically promotes thermalization \cite{Abanin2015,Mori2016,KUWAHARA201696}(with notable exceptions such as in many-body localization \cite{pal_many-body_2010,ponte_many-body_2015,Khemani2016,Else2016,Yao2017,
alet_many-body_2018,abanin_colloquium_2019,sierant_stability_2023}). The study of driven quantum many-body systems is further motivated by Floquet dynamics being naturally realized in quantum computers, where gates are discrete \cite{mi_time-crystalline_2022,zhang_many-body_2023}. Examples of such Floquet scarring are, however, lacking.

Here, we show the first instance of genuine scarring in a many-body system under a periodic drive. We focus on the mixed-field Floquet quantum Ising model, a paradigmatic model for quantum chaos \cite{dalessio_long-time_2014,ponte_many-body_2015,shukla_diagnosing_2025,staszewski_krylov_2025}.
On the one hand, scarring analogous to the static case persists in the presence of the drive, the two being related by the high-frequency limit. On the other hand, the drive can lead to additional and distinctively new forms of scarring without static counterparts. The interplay between the frequencies of drive and periodic orbits gives rise to a rich phase diagram, delineating regimes where scarring is enhanced or quenched [Fig.\ \ref{fig:phase_diag}(c,d)]. Remarkably, these regions are located by a classical analysis of the corresponding Lyapunov exponents. The transitions manifest in experimentally accessible dynamical observables, providing guidance for their detection in modern quantum devices.

{\em Model.---} We consider discrete dynamics $\ket{\psi(t)}=(U_F)^t\ket{\psi(0)}$ with integer times $t$ under the following Floquet unitary for a mixed-field Ising chain
\begin{align}
    U_F = 
    e^{-i\frac{T}{2} h \sum_j X_j}
    e^{-i\frac{T}{2}\left( J \sum_j Z_j Z_{j+1} +  h \sum_j Z_j \right)},
    \label{eq:Floquet operator}
\end{align}
where $X_j$ and $Z_j$ are Pauli-operators at sites $j=1,\dots,N$, and periodic boundary conditions $Z_{1}=Z_{N+1}$ are assumed. We fix the Ising coupling $J=1$ and tune the driving period $T$ (associated frequency {\color{blue}$\Omega=\frac{2\pi}{T}$}) and field $h$ (with equal transverse and longitudinal components). The classical spin dynamics associated to Eq.~\eqref{eq:Floquet operator} is stroboscopic \cite{Howell2019}
\begin{align}
        \bm{s}_j&(t+1)=\mathbf{R}_x[hT]\mathbf{R}_z\big[\big(h+h^{\mathrm{ex}}_j(t)\big)T\big]
    \bm{s}_j(t),
    \label{eq:classical Floquet}
\end{align}
where $\mathbf{R}_{x,z}(\theta)$ are rotation matrices by an angle $\theta$ around $x$ and $z$, respectively, and with classical exchange field $h^{\mathrm{ex}}_j(t)= J\left(s^z_{j-1}(t)+s^z_{j+1}(t)\right)$.

Except for fine-tuned parameters, this model is generally chaotic. For generic random (GR) product states, $\ket{\mathrm{GR}}=\ket{\mathbf{s}_1,\mathbf{s}_2,\dots}$ with random Bloch vectors $\mathbf{s}_i$, the classical dynamics is aperiodic and unstable (with a positive Lyapunov exponent), whereas the quantum dynamics leads to a quick relaxation of local observables. For special interaction-suppressing (IS) states, $\ket{\mathrm{IS}}=\ket{+\mathbf{s},+\mathbf{s},-\mathbf{s},-\mathbf{s},+\mathbf{s},+\mathbf{s},\dots}$, designed so that all the classical exchange fields $h^{\mathrm{ex}}_j(t)$ vanish, the classical dynamics simplifies to $\bm{s}_j(t)=\mathbf{Q}^t \bm{s}_j(0)$ with $\mathbf{Q}=\mathbf{R}_x(h T)\mathbf{R}_z(h T)$. The motion thus reduces to a simple global spin precession with period $T_s=\frac{2\pi}{\omega_s}$ and frequency $\omega_s$ determined by $\cos\frac{\omega_s T}{2}=\cos^2\frac{hT}{2}$ [Fig.\ \ref{fig:phase_diag}(a)]. In the high-frequency limit $T\to 0$, which corresponds to the non-driven case, the precession becomes continuous. This precession is, however, generically unstable to perturbations [Fig.\ \ref{fig:phase_diag}(b)]. In the quantum dynamics from the IS states, local observables can undergo a few oscillations before ultimately relaxing.

\textit{Classical Lyapunov stability.---} Quantum scarring can emerge when the classical dynamics includes short enough UPOs. Here, IS states play this role. Their classical instability is controlled by the Lyapunov exponent $\lambda_s$, and can be diagnosed by the decorrelator $d(t) = \sqrt{1-\langle \big| \langle \bm{s}_j(t) \rangle_{\mathrm{runs}} \big|^2\rangle_j}$, where $\langle \cdots \rangle_{\mathrm{runs}}$ denotes an average over an ensemble of slightly perturbed initial conditions and $\langle \cdots \rangle_j$ an average over sites. For a UPO, the decorrelator grows exponentially, $d(t) \sim d(0) e^{\lambda_s tT}$, until saturation is reached. 

A common criterion for scarring estimates the instability required for a trajectory initially close to a periodic orbit to deviate significantly over one period, or $\lambda_s\ll \omega_s$~\cite{kaplan_scars_1999}. In the high-frequency limit, we find $\lambda_s/\omega_s\propto J/h$, consistent with analytical results in the time-independent problem~\cite{pizzi_genuine_2025}. We refer to this case as the $0$-scar regime, since these scars develop for {\color{blue}$\omega_s/\Omega \simeq 0$}. Periodic driving introduces an additional timescale {\color{blue}$\Omega^{-1}$}, and scarring thus requires both $\lambda_s\ll \omega_s$ and {\color{blue}$\lambda_s\ll \Omega$}. The latter condition translates to {\color{blue}$\lambda_s/\Omega\propto JT$}, valid for small $T$, and suggests that scarring is quenched with increasing driving period. In the vicinity of $\omega_s\simeq \pi/T$, however, the IS state is periodically flipped and returns to itself only after two driving cycles [Fig.\ \ref{fig:phase_diag}(b)]. This defines a second, distinct $\pi$-scar regime. In a rotating-frame description, the relaxation time becomes comparable to that near {\color{blue}$\omega_s/\Omega \simeq 0$} when observed at even multiples of the period. For $JT \ll 1$, we estimate $\lambda_s/\omega_s \propto J/h_\pi$, with detuned field $h_\pi = h - \pi/T$. 

The stability beyond these limits is captured by our analytical calculation [details in {\color{blue}End Matter}~\ref{end matter:analytics Lyapunov}]. For 0-scars, we work in the limit {\color{blue}$\omega_s/\Omega \simeq 0$} in which the classical dynamics yields a slow precession over many drive periods. In a rotating frame, the spins are subject to an effective time-averaged interaction $\overline{\mathbf{J}}=\frac{1}{T_s}\sum^{T_s-1}_{t=0}\mathbf{Q}^{-t}
\left(J\mathbf{\hat{z}}\otimes \mathbf{\hat{z}}\right) \mathbf{Q}^t$. We linearize the classical dynamics in the small deviations $\bm{\epsilon}_j$ around the IS state, $\mathbf{s}_j = (-1)^{\lfloor \frac{j}{2} \rfloor}\mathbf{s} + \bm{\epsilon}_j$, yielding 
\begin{align}
\bm{\epsilon}_j(t+1)=\bm{\epsilon}_j(t)+(-1)^{\lfloor  \frac{j}{2}\rfloor}  \mathbf{s} \times\bigg[T\overline{\mathbf{J}}\big(\bm{\epsilon}_{j-1}(t)+\bm{\epsilon}_{j+1}(t)\big)\bigg].
\end{align}
Exploiting translation symmetry, we can solve the associated eigenproblem in momentum space. We obtain the Lyapunov exponent 
\begin{align}
\lambda_{0}
=
\frac{1}{T}\ln(1+JT g_{0}),
\label{eq:Lyapunov Ising}
\end{align}
 with the geometric factor $g^2_{0}=\frac{1}{4}\textcolor{blue}{(1-q_z^2)}[(1-3q_z^2)(\mathbf{q}\cdot \mathbf{s})^2 + 2q_z^2]$ determined by the rotation axis $\mathbf{q}$ of  $\mathbf{Q}$. Due to the periodicity of the drive, $g_0$ (and thus $\lambda_0$) depends periodically on $hT$ [explicit expressions in {\color{blue}End Matter}~\ref{end matter:analytics Lyapunov}]. 

{\color{blue}We use the classical scarring indicator $\Lambda=\max(\delta,J)/\lambda_s$, where $\delta$ is the precession rate measured from the nearest drive resonance, i.e.\ $\delta=\omega_s$ near $0$-scars and $\delta=\pi/T-\omega_s$ near $\pi$-scars; close to a resonance, $J$ is the remaining frequency scale. This generalizes the single-particle result that the scar strength grows with the ratio of orbit frequency and Lyapunov exponent~\cite{kaplan_measuring_1999} to many-body Floquet systems. Large $\Lambda$ indicates scarring: at high frequency, $\Lambda\simeq4h/J$ predicts $0$-scars for $h\gtrsim J$. Near the drive resonances $hT\simeq n\pi$, the orbit becomes marginally stable, predicting scarred corridors. In between, scarring should be quenched. We mark the crossover by a threshold $\Lambda=\Lambda_c=6$ which we use throughout  [Fig.\ \ref{fig:phase_diag}(c), App.~\ref{Lyapunov exponent: numerics vs analytics}].}

\begin{figure}[t!]
    \centering
\includegraphics[width=\linewidth]{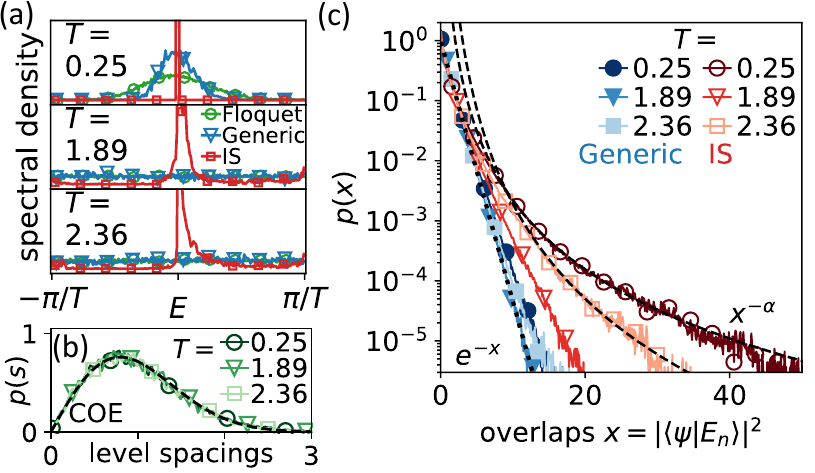}
    \caption{\textbf{Quantum statistics.} (a) Spectral densities of Floquet, IS and GR states. The Floquet spectrum broadens with increasing $T$ (green). This reflects into GR states (blue). IS states (red) cluster near $E=0$ ($T=0.25$) or near $E=0,\pi$ ($T=1.89,2.36$). (b) Level spacings follow random matrix theory (RMT). (c) Overlap distributions. GR states (blue) closely follow the Porter–Thomas form $\sim e^{-x}$. IS states (red) have fat tails, signaling the presence of scarring. We fit them with a power law $\sim x^{-\alpha}$, with $\alpha\approx 3.83$ ($T=0.25$) and  $\alpha\approx 5.25$ ($T=2.36$). At $T=1.89$, a strongly diminished tail approaches the RMT limit and signals the absence of scars. Parameters: $N=20$, $h=0.9$, $\mathcal{N}=10^{3}$.}
    \label{fig:quant_statistics}
\end{figure}

\textit{Quantum statistics.---} 
Having established the stability of the classical dynamics across parameter space, we examine how this translates into scarring of the quantum many-body spectrum. We obtain Floquet states $\ket{E_n}$, such that $U_F\ket{E_n}=e^{-iE_n T}\ket{E_n}$, by exact diagonalization in the zero-momentum, even-inversion symmetry sector. We restrict to the central $50\%$ of ordered quasi-energies $E_n\in (-\pi/T,\pi/T]$ to avoid edge effects at small $T$ where the spectrum does not span the entire Floquet circle. For $T\ll 1$, IS states cluster near $E=0$ due to $\mel{\mathrm{IS}}{U_F}{\mathrm{IS}}\approx 1$ [Fig.\ \ref{fig:quant_statistics} (a)]. Increasing $T$ then broadens the spectrum, reflected in an analogous behavior in the spectral density of GR states. This induces an additional peak in the density of IS states at $E=\pi/T$. We confirm that the system remains quantum chaotic across these driving periods: the distribution of consecutive level-spacings $\delta_n=(E_{n+1}-E_n)/\delta $ (mean-level spacing $\delta$)  follows the circular orthogonal ensemble of random matrix theory (RMT) [Fig.\ \ref{fig:quant_statistics} (b)]. 

To probe scarring, we analyze the statistics of overlaps $x_{n}=\mathcal{D}|\mel{E_n}P{\psi}|^2$. The projector $P$ (with $P^2=P$) onto the symmetry sector of Hilbert-space dimension $\mathcal{D}$  ensures that the less-symmetric GR states are treated on equal footing with IS states \cite{pizzi_genuine_2025}. We sample $|\psi\rangle$ from GR and IS states uniformly in their respective Hilbert spaces. The overlaps of GR states closely follow the Porter–Thomas distribution, $p_\mathrm{PT}(x)=e^{-x}$, for all considered $T$ [Fig.\ \ref{fig:quant_statistics} (c)], consistent with the expectation that chaotic Floquet eigenstates behave as Haar-random vectors.

By contrast, the overlap distribution for IS states depends strongly on the driving period. For small $T$, in the 0-scar regime, we find a pronounced fat tail at large $x$, well fit by a non-universal power-law, signaling anomalously large IS weight and thus scarring. At intermediate $T$, the tail is drastically suppressed and approaches the Porter-Thomas limit. Finally, for larger $T$ in the $\pi$-scar regime, a power-law behavior reemerges. The driving period thus acts as a tuning knob to access various regimes of scarring. 

\begin{figure}[t!]
    \centering
\includegraphics[width=\linewidth]{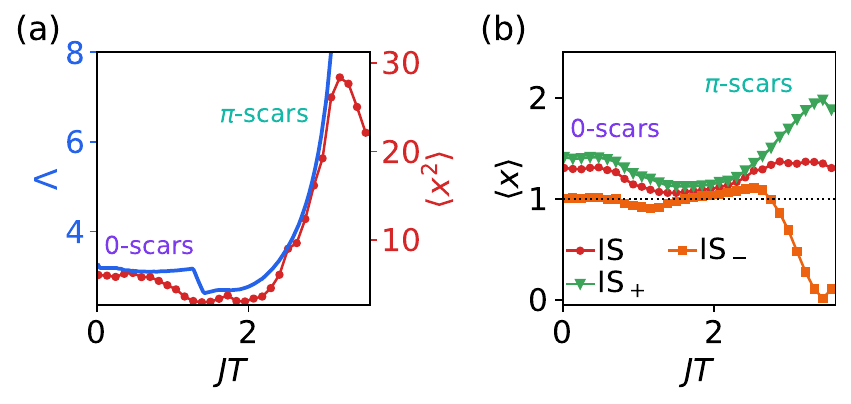}
    \caption{
    {\color{blue}
    \textbf{Lyapunov exponent vs.~scarring.} (a) Classical scarring indicator $\Lambda$ (blue solid) and second overlap moment $\langle x^2\rangle$ of IS states \textcolor{blue}{(red circles)}. (b) Mean overlap for IS states and their coherent spin-flipped superpositions ($\mathrm{IS}_\pm$), \textcolor{blue}{indicating} enhanced overlaps consistent with the semiclassical analysis. Parameters: $N=16$, $h=0.9J$.
    }
    }
    \label{fig:lyapunov_vs_scarring}
\end{figure}

\begin{figure*}[t!]
    \centering
\includegraphics[width=\linewidth]{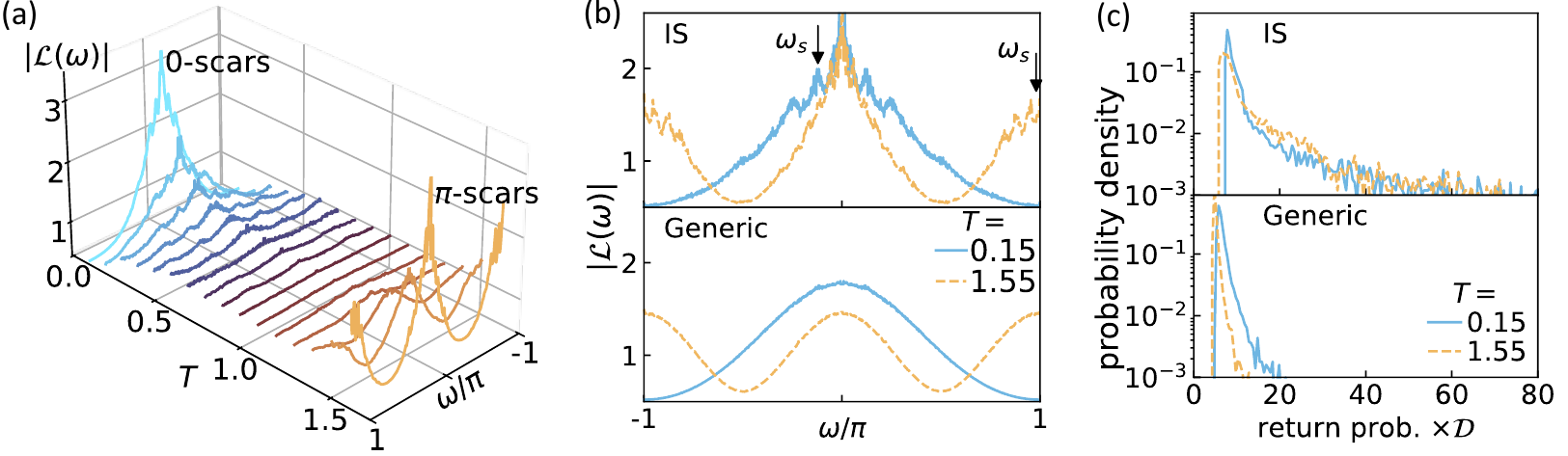}
    \caption{\textbf{Quantum dynamics.}
    (a) Fourier transform of the average Loschmidt echo for IS states {\color{blue}(colors for contrast correspond to different $T$)}. Scars yield resonances at integer multiples of $\omega_s$, prominent near $T\simeq 0$ (light-blue) and $T\simeq 1.5$ (orange). For intermediate $T$, they vanish. (b) Top: Same as (a) at fixed $T$ (see colors). $0$-scars have a band of peaks near $\omega= 0$. $\pi$-scars exhibit odd-integer $\omega_s$ peaks near $\omega=\pi$ and an even-integer band near $\omega=0$ from backfolding. Bottom: GR states are featureless, indicating no long-time memory. (c) Distribution of time-averaged rescaled return probabilities. IS states are broad with enhanced revivals, while GR states cluster around $\approx 3.2$. Parameters: $N=16$, $h_{x,z}=1.8$, (a) $\mathcal{N}=10^{3}$, (b,c) $\mathcal{N}=10^{4}$.}
    \label{fig:quant_dyn}
\end{figure*}

\textit{Stability diagram.---}
Next, we map the effectiveness of scarring across parameter space by computing the moments of the overlap distribution in the $(h/J,JT)$ plane.  For the IS states, the resulting landscape reveals stripe-like corridors where scarring is either enhanced or quenched  [Fig.\ \ref{fig:phase_diag} (c)]. The corridors are characterized by anomalously large moments  $\expval{x}\approx 1.4$ and $\langle x^2\rangle \approx 10$, far exceeding the RMT prediction $\langle x_{\mathrm{PT}} \rangle=1$ and $\langle x_{\mathrm{PT}}^2 \rangle=2$. Here, $\expval{\dots}$ is the combined spectral and state average. Between these corridors, the moments approach the Porter-Thomas values such that scarring effectively disappears. 

Figure \ref{fig:phase_diag} (c) {\color{blue} and Fig.~\ref{fig:lyapunov_vs_scarring}} compare the classical scarring indicator $\Lambda$ \textcolor{blue}{with the second overlap moment $\langle x^2\rangle$. Here, \textcolor{blue}{$\langle x^2\rangle$} plays the role of the inverse participation ratio (IPR) \textcolor{blue}{and $\Lambda$ that of $1/\lambda$} in the single-particle relation $\mathrm{IPR}\propto1/\lambda$~\cite{kaplan_measuring_1999}. The crossover $\Lambda=\Lambda_c$ follows the boundaries of the scarred regions. We note that the quantum corridors are wider than the classical resonance windows of width $\sim JT$. Along a cut at fixed field, $\Lambda$ tracks $\langle x^2\rangle$ quantitatively [Fig.\ \ref{fig:lyapunov_vs_scarring}(a)]. Thus, we confirm that the classical dynamics dictates the properties of the quantum many-body wavefunction.} 

The stripe-like corridors correspond to the two scar families and can also be understood quantum mechanically. The corridor centered at $\omega_s=0$ hosts 0-scars. They smoothly connect to the static case and are associated to the effective Hamiltonian $H_{\mathrm{eff}} (h) = \frac{h}{2} \sum_j (X_j+Z_j)+\frac{J}{2} \sum_j Z_j Z_{j+1}$ in the high-frequency limit. {\color{blue}While $H_{\mathrm{eff}}$ describes the system at short times, scarring remains stable at long times despite Floquet heating: $0$- and $\pi$-scars exist where the Floquet heating of the IS state is incomplete, described by a diagonal ensemble at lower effective temperature than the infinite-temperature state [App.~\ref{app:heating}], whereas the intermediate, unscarred corridor heats up completely.}

The corridor centered at {\color{blue}$\omega_s=\pi/T$} hosts $\pi$-scars. These appear when the field adds a $\pi$-flip, with operator $D=\prod_j Y_j$. For $JT\ll 1$, the associated Floquet operator can be written as $U_\pi \approx De^{-iH_\pi T} $ with effective Hamiltonian $H_\pi=H_{\mathrm{eff}}(h=h_\pi)$. 
{\color{blue} While the rotating
frame naturally accounts for the $\pi$-scars by an effective Hamiltonian identical in structure to the high-frequency
case, $0$- and $\pi$-scar regime are not smoothly connected. To probe the non-equilibrium nature of the $\pi$-scars, we consider the overlap with the cat-states $\ket{\mathrm{IS}}_\pm=(\ket{\mathrm{IS}}\pm\ket{\mathrm{IS}'})/\sqrt{2}$ where $\ket{\mathrm{IS}'}=D\ket{\mathrm{IS}}$ denotes the spin-flipped IS partner. Near the $\pi$-scar regime at $JT\simeq3.4$, $\langle x\rangle_{\mathrm{IS}_+}\simeq2.0$ exceeds the already-enhanced plain-IS overlap $\langle x\rangle_{\mathrm{IS}}\simeq1.4$, while $\langle x\rangle_{\mathrm{IS}_-}\simeq0.01$ is suppressed to almost zero [Fig.\ \ref{fig:lyapunov_vs_scarring} (b)]. We find these features remain stable under system-size scaling [App.~\ref{app:system-size scaling}]. This coherent splitting shows that the $\pi$-flip leaves a genuine imprint on the many-body wavefunction rather than merely relabeling the $0$-scar in a rotating frame.}

We monitor the level spacing ratio $r=\min(\delta_n,\delta_{n+1})/\max(\delta_n,\delta_{n+1})$  to rule out an integrable origin of these structures [End Matter, Fig.\ \ref{fig:end_matter_phase_diag}]. The system remains mostly chaotic with $r\approx 0.53$, except for a narrow quasi-integrable strip with Poissonian $r\approx 0.39$ near $JT=\pi$. This regime emerges as the Ising unitary approaches the identity (up to a global phase), $e^{-i\frac{\pi}{2}\sum_j Z_jZ_{j+1}}\propto \mathbf{1}$, reducing the dynamics to pure field terms. The overlap moments of GR states closely track the RMT predictions, $\expval{ x_{\mathrm{PT}}}=1$, across the  parameter plane with small deviations at the percent-level [Fig.\ \ref{fig:phase_diag} (d)]. Deviations appear in the second moment $\langle  x_{\mathrm{PT}}^2\rangle=2$ near $JT= \pi$ [End Matter, Fig.\ \ref{fig:end_matter_phase_diag}]. {\color{blue} The overlap enhancement in the scarred corridors persists under system-size scaling  [Fig.\ \ref{fig:phase_diag}(e)].  The transition between scarred and unscarred regimes remains a smooth crossover for $N\to \infty$.}

\textit{Quantum dynamics.---}
Finally, we turn to the dynamical signatures of Floquet scars. We probe the Loschmidt echo
\begin{align}
    \mathcal{L}(t) = \left|\braket{\psi(0)}{\psi(t)}\right|^2=
    \frac{1}{\mathcal{D}^2}\sum_{nm}
    x_nx_m e^{-i(E_n-E_m)t}
\label{eq:Loschmidt}
\end{align}
for the GR and IS families of symmetry-projected initial states. 
This can be accessed experimentally in quantum platforms, e.g., in trapped ions \cite{jurcevic_direct_2017} or superconducting qubits \cite{braumuller_probing_2022}. 

GR states display a featureless Fourier spectrum $\mathcal{L}(\omega)$ [Fig.\ \ref{fig:quant_dyn} (b)], signaling fully relaxed long-time dynamics and thus no long-time memory of their initial conditions. In contrast, for IS states, the Loschmidt echo is enhanced because of the anomalously large overlaps. Scarring manifests as characteristic resonances at integer multiples of $\omega_s$ in  $\mathcal{L}(\omega)$. For $0$-scars, we find a band of peaks near $\omega=0$. For $\pi$-scars, the primary peaks near $\omega=\pi$ for odd integer $k$ are back-folded at even integers at {\color{blue}$\omega_s/\Omega\simeq 0$}. This behavior reflects many terms $E_n-E_m\simeq k \omega_s$ with integer $k$ in the sum in Eq.\ \eqref{eq:Loschmidt}. 
We further observe weaker replica peaks at $\omega\simeq k\omega_s+2lJT$ with integer $l$ for $h_\pi\ll J$, which we attribute to magnon excitations from the IS state [End Matter \ref{end matter: Loschmidt echo}]; these features diminish for $h_\pi>J$.

We track the evolution of these peaks as a function $T$ [Fig.\ \ref{fig:quant_dyn} (a)].  First, they progressively smear out and vanish as a function of $T$. As one approaches the $\pi$-scar regime, the  dynamic signatures are recovered, consistent with the spectral findings above. This demonstrates that the scarring transition is accessible in dynamic observables and complements the statistical perspective.

Scars also heavily impact the distribution of long-time return probabilities, $p(|\mathcal{L}_{\omega=0}|)$ [Fig.\ \ref{fig:quant_dyn} (c)]. For GR states, this distribution is sharply peaked, while IS states produce a fat tail. This reflects their distinct overlap statistics: the return probability is the sum in Eq.\ \eqref{eq:Loschmidt}, restricted to $E_n-E_m\simeq 0$, over random numbers $x_n$, which we assume to be statistically independent. For IS states, the sum of fat-tailed terms 
yields, in turn, a power-law return probability, consistent with a Lévy stable law \cite{bouchaud_anomalous_1990,schmid_robust_2024}. For GR states, the narrow overlap statistics leads to a normal distribution via the conventional central limit theorem. Note that Fig.\ \ref{fig:quant_dyn}(c) shows $p(|\mathcal{L}_{\omega=0}|)$ for convenience rather than $p(\mathcal{L}_{\omega=0})$, which modifies the Gaussian.

\textit{Conclusions.---} We have demonstrated genuine quantum scarring in Floquet many-body spin systems. Scarring derived from IS spin configurations manifests both in quantum statistics and dynamics, as diagnosed in enhanced overlaps with Floquet eigenstates and by the Loschmidt echo. The interplay between the frequency of the periodic orbit and the drive gives rise to a rich spectrum of behaviors, including absence of scarring, $0$-scars, and $\pi$-scars, identifying the drive period as a key tuning knob. We have verified the phenomenology in the mixed-field Floquet quantum Ising model as well as Floquet XXZ chains [End Matter \ref{end matter: XXZ}], suggesting that genuine Floquet scarring is a generic feature of chaotic spin systems.

Our work opens several avenues for future investigation. Beyond $0$-scars and $\pi$-scars, one may ask whether higher-order scars exist. A natural testbed are long-range systems that host higher-order discrete time crystals \cite{pizzi_higher-order_2021,giachetti_fractal_2023,liu_higher-order_2024,Roberto2024}. Another open question regards the role of higher order terms in the Magnus expansion: while IS states give UPOs at zeroth order, can other terms support distinct UPOs, and how does this affect scarring?
It is also natural to explore scarring-like effects in systems under structured random \cite{zhao_random_2021,zhao_temporal_2023,liu_prethermalization_2026} or quasi-periodic drives \cite{dumitrescu_logarithmically_2018,peng_time-quasiperiodic_2018,crowley_topological_2019,lapierre_fine_2020,schmid_self-similar_2025}.
Finally, it will be highly desirable to observe genuine Floquet scars in quantum simulator platforms.

\textit{Data availability.---} The code is available on Zenodo \cite{zenodo}. 

\textit{Use of AI.---} Claude Code (Opus 5.5) was used to assist with algebraic checks, code development, and language editing. All AI-assisted material was reviewed and revised by the authors, who take full responsibility for the content.

\begin{acknowledgments}
\textit{Acknowledgments.---}
We thank C.~Castelnovo, C.~B.~Dag, B.~Evrard, and L.~H.~Kwan for many insightful discussions and collaborations on related topics. A.~P acknowledges support from Trinity College Cambridge. We acknowledge the use of QuSpin \cite{weinberg_quspin_2017,weinberg_quspin_2019}. JK acknowledges support from the Deutsche Forschungsgemeinschaft (DFG, German Research Foundation) under Germany’s Excellence Strategy–EXC– 2111–390814868, DFG grants No. KN1254/1-2, KN1254/2-1, and TRR 360 - 492547816, as well as the Munich Quantum Valley, which is supported by the Bavarian state government with funds from the Hightech Agenda Bayern Plus.
\end{acknowledgments}

\vfill


\bibliographystyle{apsrev4-1}
\bibliography{bibliography}

\clearpage
\renewcommand{\appendixname}{End Matter} 
\appendix
\onecolumngrid
\begin{center}
\includegraphics[width=0.85\linewidth]{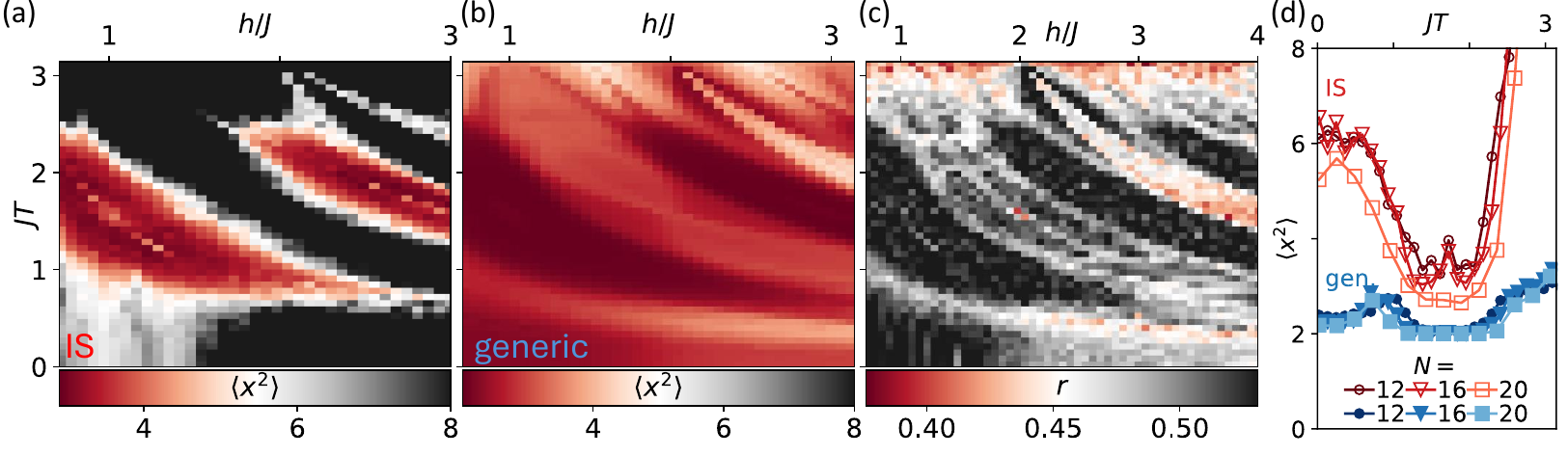}
\makeatletter\def\@captype{figure}\makeatother
\caption{Second overlap moment $\langle x^2 \rangle$. (a) IS states [same as Fig.\ \ref{fig:phase_diag} (c)]. (b) GR states remain mostly close to the RMT prediction  $\langle x^2_\mathrm{PT}\rangle =2$. (c) Level spacing ratio indicates mostly chaotic behavior with $r=r_\mathrm{COE}=0.531$, consistent with RMT. Quasi-integrable regions with  $r \simeq r_\mathrm{Poisson}=0.386 $  coincide with pronounced features for GR states. (d) Finite-size scaling at $h=0.9J$. Same parameters as in Fig.\ \ref{fig:phase_diag}.  }
\label{fig:end_matter_phase_diag}
\end{center}
\twocolumngrid
\begin{figure*}[t!]
    \centering
\includegraphics[width=0.85\linewidth]{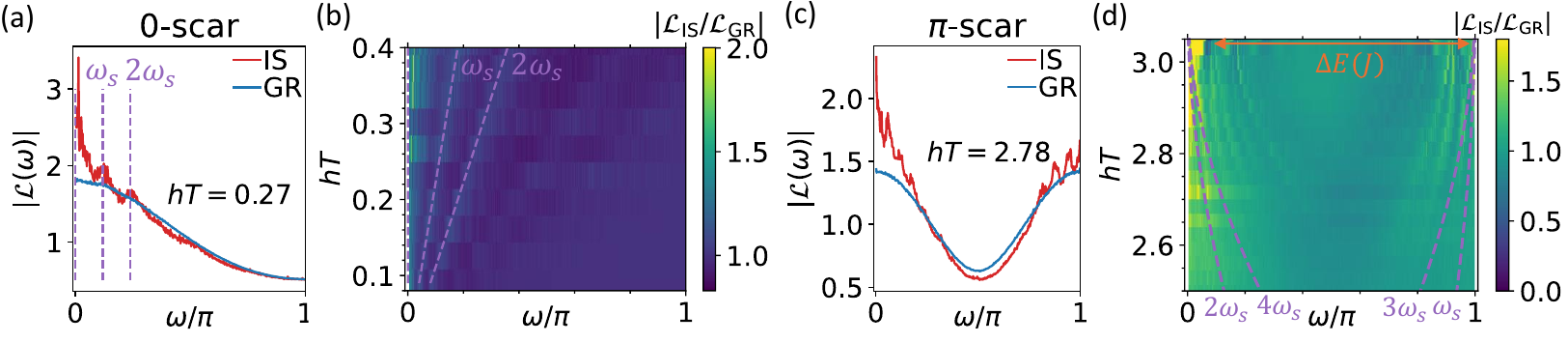}
    \caption{Resonances in the Loschmidt echo. (a,b) $0$-scars. (b) After normalization by generic states, the echo shows approximately linear dispersing resonances with $hT$. (c,d) $\pi$-scars. (d) Resonances disperse approximately quadratically with $hT$ near $hT\simeq \pi$; the scar frequency $\omega_s$ serves as a guide to the eye. For $h_\pi< J$ near $hT\simeq \pi$, additional magnon replicas appear at $\Delta E(J) \simeq 2JTl$ ($l \in \mathbb{N}$) which diminish for  $h_\pi> J$. Same parameters as in Fig.\ \ref{fig:quant_dyn}.}
\label{fig:Loschmidt echos}
\end{figure*}
\section{Analytical Lyapunov exponent} 
\label{end matter:analytics Lyapunov}

We analytically derive the Lyapunov exponent for IS states $\mathbf{s}_j=(-1)^{\lfloor \frac{j}{2}\rfloor}\mathbf{s}$, in analogy with the time-independent case \cite{pizzi_genuine_2025} but accounting for the drive. \textcolor{blue}{The derivation averages the exchange dynamics over the precession of the IS orbit and therefore requires this precession, measured from the nearest drive resonance, to be fast on the scale of the exchange coupling. The resonances are $\omega_sT\simeq0$ ($0$-scars) and $\omega_sT\simeq\pi$ ($\pi$-scars), i.e.\ $hT\simeq n\pi$.}

We transform into the rotating frame of the scar
\begin{align}
\mathbf{u}_j(t)=\left(\mathbf{Q}^{T}\right)^t\mathbf{s}_j(t),
\end{align}
where $\mathbf{Q}=\mathbf{R}_x(h T)\mathbf{R}_z(h T)$. The dynamics becomes
\begin{align}
    \mathbf{u}_j(t+1)=
    \mathbf{\tilde{R}}_{zz}(j,t) \mathbf{u}_j(t)
\end{align}
where the bare interaction $\mathbf{R}_{zz}(j,t)=\mathbf{R}_z\big[h^{\mathrm{ex}}_j(t) T\big]$ in the rotating frame is
$\mathbf{\tilde{R}}_{zz}(j,t) = 
    \left(\mathbf{Q}^{T}\right)^{t+1} \mathbf{R}_{zz}(j,t) (\mathbf{Q})^{t+1}$.

We average the transformed interaction over the fast degrees of freedom, 
\begin{align}
    \mathbf{\tilde{R}}_{zz}(j,t)\rightarrow \overline{\mathbf{\tilde{R}}_{zz}(j,t)}=\frac{1}{T_s}\sum^{T_s-1}_{t=0}\mathbf{\tilde{R}}_{zz}(j,t).
\end{align}
To take powers $(\mathbf{Q})^t$, we consider the rotation axis $\mathbf{q}$ and angle $\omega_s$ and use Rodrigues' formula
\begin{align}
    (\mathbf{Q})^t= \mathbf{q} \otimes \mathbf{q} +\sin(\omega_s t T) [\mathbf{q}]_\times + \cos(\omega_s t T)(\mathbf{1}-\mathbf{q} \otimes \mathbf{q}),
\end{align}
where $[\mathbf{q}]_\times$ is the cross-product matrix, such that $\bm{q} \times \bm{a} = [\mathbf{q}]_\times \bm{a}$. Averaging yields an effective interaction
\begin{align}
     \overline{\mathbf{\tilde{R}}_{zz}(j,t)}
     &=(\mathbf{q} \otimes \mathbf{q})\mathbf{R}_{zz}(j,t)
     (\mathbf{q} \otimes \mathbf{q})
     -\frac{1}{2}[\mathbf{q}]_\times \mathbf{R}_{zz}(j,t)
     [\mathbf{q}]_\times
     \notag
     \\
     &+\frac{1}{2} (\mathbf{1}-\mathbf{q} \otimes \mathbf{q})
     \mathbf{R}_{zz}(j,t)(\mathbf{1}-\mathbf{q} \otimes \mathbf{q}).
\end{align}
In the regime of interest, the perturbed trajectory is close to an IS state, and the exchange field $h^{\mathrm{ex}}_j(t)$ is small, justifying an expansion $\mathbf{R}_{zz}(j,t)\simeq \mathbf{1}-iT\mathbf{J}\left(\mathbf{s}_{j+1}(t)+\mathbf{s}_{j-1}(t)\right)$. We have defined the matrix $\mathbf{J}=J\mathbf{\hat{z}} \otimes \mathbf{\hat{z}}$ with the temporal average $\overline{\mathbf{J}}=\frac{1}{T_s}\sum^{T_s-1}_{t=0} \left(\mathbf{Q}^{T}\right)^{t+1} \mathbf{J} (\mathbf{Q})^{t+1}$\textcolor{blue}{$=J\left[q_z^2\,\mathbf{q}\otimes\mathbf{q}+\frac{1}{2}(1-q_z^2)(\mathbf{1}-\mathbf{q}\otimes\mathbf{q})\right]$}.

Next, we linearize the perturbed dynamics in the small deviations $\bm{\epsilon}_j(t)$ from the IS state, $\mathbf{s}_j = (-1)^{\lfloor \frac{j}{2} \rfloor}\mathbf{s} + \bm{\epsilon}_j$,
\begin{align}
\bm{\epsilon}_j(t+1)&=\bm{\epsilon}_j(t)+(-1)^{\lfloor  \frac{j}{2}\rfloor} T \mathbf{s} \times\big[\overline{\mathbf{J}}(\bm{\epsilon}_{j+1}(t)+\bm{\epsilon}_{j-1}(t))\big]
\end{align}
In momentum space,   
$\bm{\epsilon}_k(t)=\frac{1}{\sqrt{N}}\sum_{j}
\bm{\epsilon}_j(t)e^{-ikj}$, we find
$\bm{\epsilon}_k(t+1)= \sum_{k^\prime}    \bm{\Lambda}_{k,k^\prime}
  \bm{\epsilon}_{k^\prime}(t)$, with stability matrix
\begin{align}
    \bm{\Lambda}_{k,k^\prime}
=
\delta_{k,k^\prime}
-
\sqrt{2}\sin(k)T [\mathbf{s}]_\times \overline{\mathbf{J}}
\sum_{\sigma=\pm}
\sigma  e^{-i\frac{\sigma \pi}{4}}
\delta_{k,k^\prime+\frac{\sigma \pi}{2}}.
\end{align}
This defines the eigenproblem $e^{T \lambda_{k,\alpha} }\bm{E}_{k,\alpha} =  \bm{\Lambda} \bm{E}_{k,\alpha}$ with eigenstates $\bm{E}_{k}=\bigl(\bm{\epsilon}_{k},\bm{\epsilon}_{k+\pi/2},\bm{\epsilon}_{k+\pi},\bm{\epsilon}_{k-\pi/2})^T$ labeled by $\alpha$. The Lyapunov exponent can be identified as $\lambda_0=\mathrm{max}_{k,\alpha}\lambda_{k,\alpha}$.

Proceeding analogous to the time-independent case \cite{pizzi_genuine_2025}, one finds for the generic Floquet case
\begin{align}
(e^{\lambda_{0}T}-1)^2=\frac{T^2}{4} &\left| \mathbf{q}^T \mathbf{J} \mathbf{q} - \mathrm{Tr}\,\mathbf{J} \right| 
\notag
\\
\times & \left|  (\mathbf{q} \cdot \mathbf{s})^2 \left( 3\, \mathbf{q}^T \mathbf{J} \mathbf{q} - \mathrm{Tr}\,\mathbf{J} \right) - 2\, \mathbf{q}^T \mathbf{J} \mathbf{q}\right|.
\end{align}
For the Ising model this becomes Eq.\ \eqref{eq:Lyapunov Ising}. To remove the dependence on the initial conditions of the geometric factor it is convenient to replace  $g_{0}\to \sqrt{\expval{g_{0}^2}_s}$ in the average $\langle \lambda_{0}\rangle$, using {\color{blue} $\langle (\mathbf{q}\cdot \mathbf{s})^2\rangle_\mathbf{s}=\frac{1}{3}$ and $q_x^2=q_z^2=\frac{1+\cos hT}{3+\cos hT}$}\textcolor{blue}{, $q_y^2=\frac{1-\cos hT}{3+\cos hT}$}. The rotation axis is $\mathbf{q}\propto \left(1+\cos(hT)\right)\left(\mathbf{\hat{x}}+\mathbf{\hat{z}}\right)+\sin(hT)\mathbf{\hat{y}}$. The periodic dependence of the Lyapunov exponent originates from its dependence on the rotation axis of the rotating frame. One finds the simple result
\begin{align}
    \textcolor{blue}{g_0\to\sqrt{\expval{g_{0}^2}_s}=\frac{\sqrt{(3+2\cos hT)/3}}{3+\cos hT}.}
\end{align}
In the limit $T\rightarrow 0$ the Lyapunov exponent {\color{blue}$\lambda_0=J\sqrt{5/48}\simeq0.32J$} becomes independent of $hT$. \textcolor{blue}{At finite $T$ the averaged exponent depends only weakly on $hT$. The averaging fails close to the resonances $hT\simeq n\pi$, where $\mathbf{Q}$ \textcolor{blue}{is the identity ($n$ even) or} a $\pi$ rotation \textcolor{blue}{($n$ odd)}. The rotating-frame dynamics then reduces to the Ising evolution, under which the IS orbit is marginally stable. The exact exponent drops to zero there within a window of width of order $JT$. Away from this window, the same averaged expression applies on both sides of $hT=\pi$ [App.~\ref{Lyapunov exponent: numerics vs analytics}].}

\section{Resonance frequencies in the Loschmidt echo} 
\label{end matter: Loschmidt echo}

We extract resonance frequencies from the Loschmidt echo Eq.\ \eqref{eq:Loschmidt} of IS states. To isolate scar-specific features, we normalize by generic states and analyze $|\mathcal{L}_{\mathrm{IS}}/\mathcal{L}_{\mathrm{GS}}|(\omega)$.

Figure \ref{fig:Loschmidt echos} shows the dispersion of the resonances vs.\ $hT$. For $0$-scars, peaks disperse linearly [Fig.\ \ref{fig:Loschmidt echos} (b)], consistent with $\omega_s \propto h$ near $\omega_s\simeq 0$. We reproduce the spectrum by including harmonics $k\omega_s$. 

For $\pi$-scars, peaks disperse quadratically [Fig.\ \ref{fig:Loschmidt echos} (d)], consistent with $\pi/T-\omega_s \propto h^2$ near $\omega_s\simeq \pi/T$. Harmonics $k\omega_s$ again capture the spectrum for $h_\pi>J$. For $h_\pi\ll J$, additional lines appear at $\textcolor{blue}{\Delta E(J)\,T}=2JTl \mod{\textcolor{blue}{\pi}}$ ($l \in \mathbb{N}$), reflecting interaction effects. At $hT=\pi$,
\begin{align}
    U_F = D e^{i\frac{\pi J}{2}\sum_{j} Z_j Z_{j+1}}.
\end{align}
where $D$ is the spin-flip operator. The many-body spectrum consists of $z$-polarized ``cat'' states $\frac{1}{\sqrt{2}}(1\pm D)\ket{\uparrow \uparrow \downarrow \uparrow \dots }$ with glassy patterns. Their quasi-energy differences are
\begin{align}
    \textcolor{blue}{(E_n - E_m)T} =\pm \pi +2lJT \mod{\pi}
\end{align}
where $l$ counts the number of magnon pairs  (or shifted by $\pi$). These excitations fade as $T$ is lowered away from $hT=\pi$ for $h_\pi>J$.  

\begin{figure}[!b]
    \centering
\includegraphics[width=0.65\linewidth]{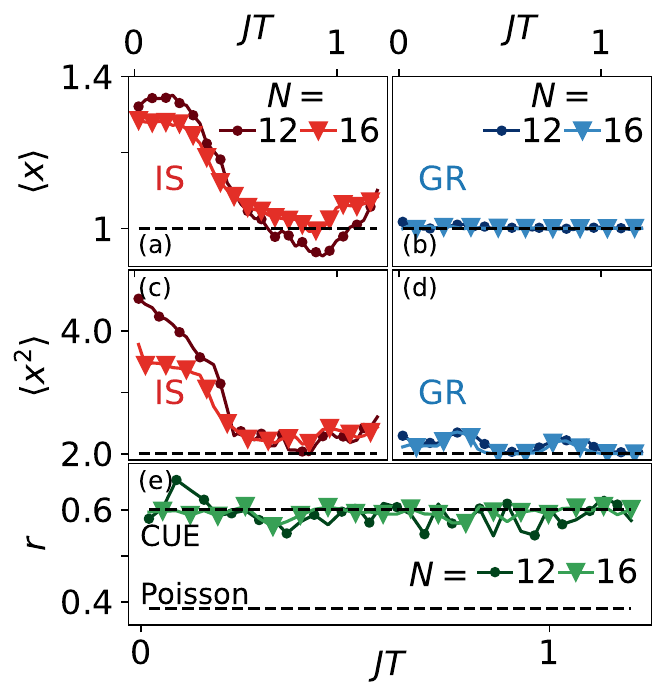}
    \caption{XXZ chain. (a,b) Mean overlap for IS and GR states. Dotted line: RMT. (c,d) Variance of the overlap for IS and GR states. Dotted line: RMT. (e) Level spacing ratio. Parameters: $J=1$, $\Delta=-0.15$, {\color{blue}$h_x=0.2$}, $h_y=-0.64$, $h_z=0.82$.}
\label{fig:XXZ moments}
\end{figure}

\section{Floquet XXZ chain with field} 
\label{end matter: XXZ}

We study scarring in the Floquet XXZ chain in a field
\begin{align}
U_F&=e^{-iT(h_x\sum^N_{j=1}X_j+J\sum^N_{j=1}X_jX_{j+1})}
\notag
\\
&\times e^{-iT(h_y\sum^N_{j=1}Y_j+J\sum^N_{j=1}Y_jY_{j+1})}
\notag
\\
&\times
e^{-iT(h_z\sum^N_{j=1}Z_j+\Delta\sum^N_{j=1}Z_jZ_{j+1})}.
\end{align}
We compute the first two overlap moments in the zero-momentum, even-inversion sector. For IS states, a scarring crossover occurs increasing $T$: from $\langle x \rangle \approx 1.3$, $\langle x^2 \rangle \approx 4$ at $T\simeq 0$ to the Porter-Thomas value   $\langle x \rangle \approx 1$, $\langle x^2 \rangle \approx 2$ at $T\simeq 1$) [Fig. \ref{fig:XXZ moments} (a,c)]. GR states stay near to Porter-Thomas [Fig. \ref{fig:XXZ moments} (b,d)]. The system remains chaotic across all $T$, consistent with CUE level statistics [Fig. \ref{fig:XXZ moments} (e)].

\clearpage
\renewcommand{\appendixname}{Appendix} 
\onecolumngrid 
\appendix
\section{Lyapunov exponent in parameter space}
\label{Lyapunov exponent: numerics vs analytics}

\begin{figure*}[t!]
    \centering
\includegraphics[width=\linewidth]{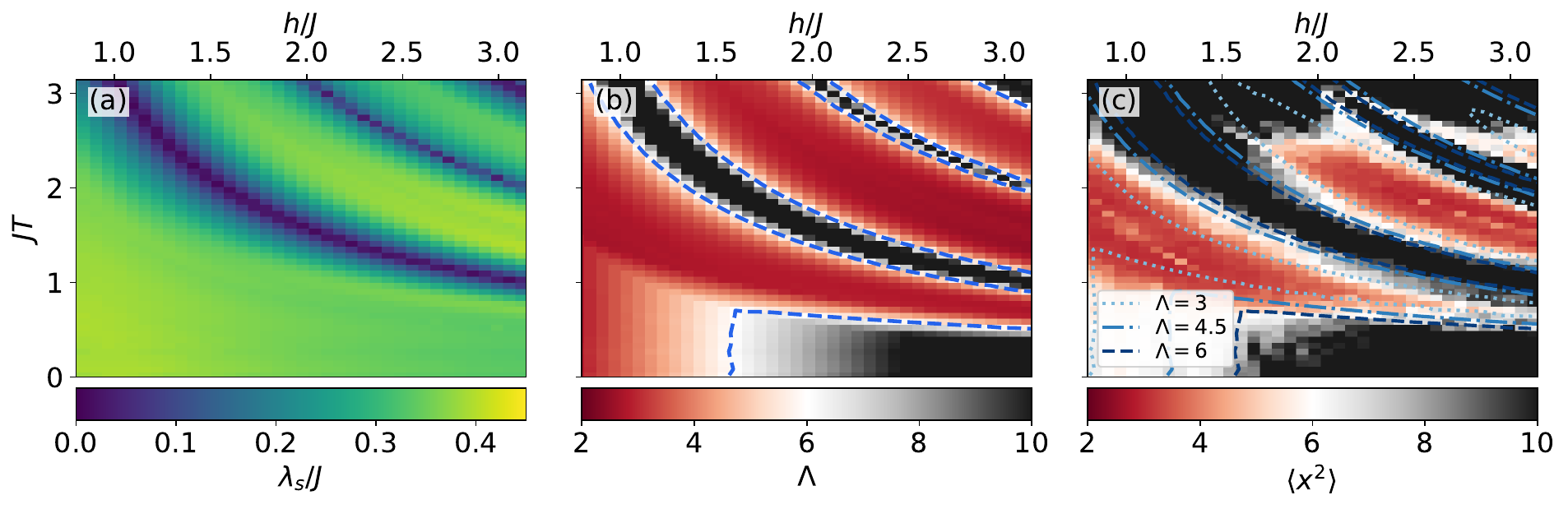}
    \caption{\textcolor{blue}{\textbf{Lyapunov exponent and scarring indicator.} (a) Maximal Lyapunov exponent $\lambda_s$ of the IS orbit from the linearized dynamics. (b) Indicator $\Lambda$. Blue dashed: $\Lambda=\Lambda_c=6$. (c) $\langle x^2\rangle$ of IS states with $\Lambda=3$ (dotted), $4.5$ (dash-dotted), and $6$ (dashed). Parameters: $N=16$.}}
\label{fig:Lyapunov}
\end{figure*}

\textcolor{blue}{We compute the maximal Lyapunov exponent $\lambda_s$ of the IS orbit from the exact linearized map of the classical dynamics [Eq.\ \eqref{eq:classical Floquet}]. We propagate the IS orbit of $N=16$ spins together with a tangent vector up to time $tT=80$ (between $300$ and $3000$ periods), and average the mean logarithmic growth rate over random IS directions $\mathbf{s}$. This is the growth rate of the decorrelator $d(t)$ for infinitesimal perturbations. Away from the drive resonances, $\lambda_s\simeq0.3J$ agrees with the averaged analytical exponent. Along $hT\simeq n\pi$ it drops to zero [Fig.\ \ref{fig:Lyapunov}(a)]. The indicator $\Lambda$ reproduces the structure of $\langle x^2\rangle$ [Fig.\ \ref{fig:Lyapunov}(b,c)]. Since the transition between scarred and unscarred regions is a crossover, we show three thresholds $\Lambda_c=3$, $4.5$ and $6$, which bracket it.}

\section{Heating}
\label{app:heating}
{\color{blue}
Quantum scarring has so far been studied in autonomous Hamiltonian systems, where the periodic orbit is confined to a fixed energy shell and the corresponding eigenstates are organized by energy; many-body ETH is likewise formulated within a microcanonical energy window. Generic Floquet dynamics does not conserve energy, so the distribution of a state over the energy sectors of $H_{\mathrm{eff}}$ is not conserved either, and the coherent recurrences responsible for scarring must survive this drive-induced redistribution. In the absence of additional conservation laws, Floquet ETH predicts heating toward the infinite-temperature ensemble; the question we address here is whether genuine many-body scars remain visible while this redistribution takes place.

To quantify the redistribution, we perform a heating analysis \cite{Abanin2015,Mori2016,KUWAHARA201696}. We compare the exact Floquet dynamics to the zeroth-order Magnus Hamiltonian, 
\begin{align}
  H_0=\frac{h}{2}\sum_j(X_j+Z_j)+\frac{J}{2}\sum_j Z_jZ_{j+1}, 
\end{align}
($H_{\mathrm{eff}}$ in the main text). All moments of $H_0$ are conserved under Hamiltonian evolution generated by $H_0$ alone. 
To diagnose heating, we consider
\begin{equation}
\overline{\mathrm{var}_{\psi_t}(H_0)}
=
\overline{
\langle H_0^2\rangle_{\psi_t}
-
\langle H_0\rangle_{\psi_t}^2
},
\end{equation}
where the overline denotes the state-ensemble average. We choose the variance instead of the mean energy density which is a poor heating diagnostic here because both the IS and GR ensembles are centered near the middle of the spectrum. Their
ensemble-averaged $\langle H_0\rangle/L$ is therefore already close to its
infinite-temperature value at the beginning of the Floquet evolution [Fig. \ref{fig:heating dynamics}(a,c)], although the individual states are not thermal.

\begin{figure*}[t!]
\centering
\includegraphics[width=0.48\linewidth]{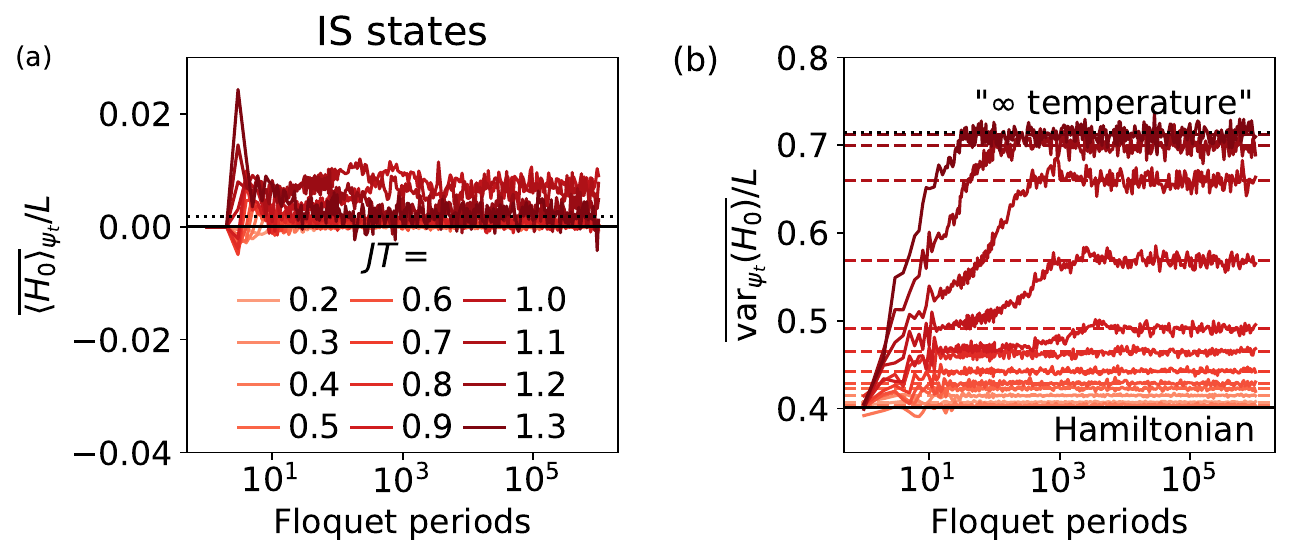}
\includegraphics[width=0.48\linewidth]{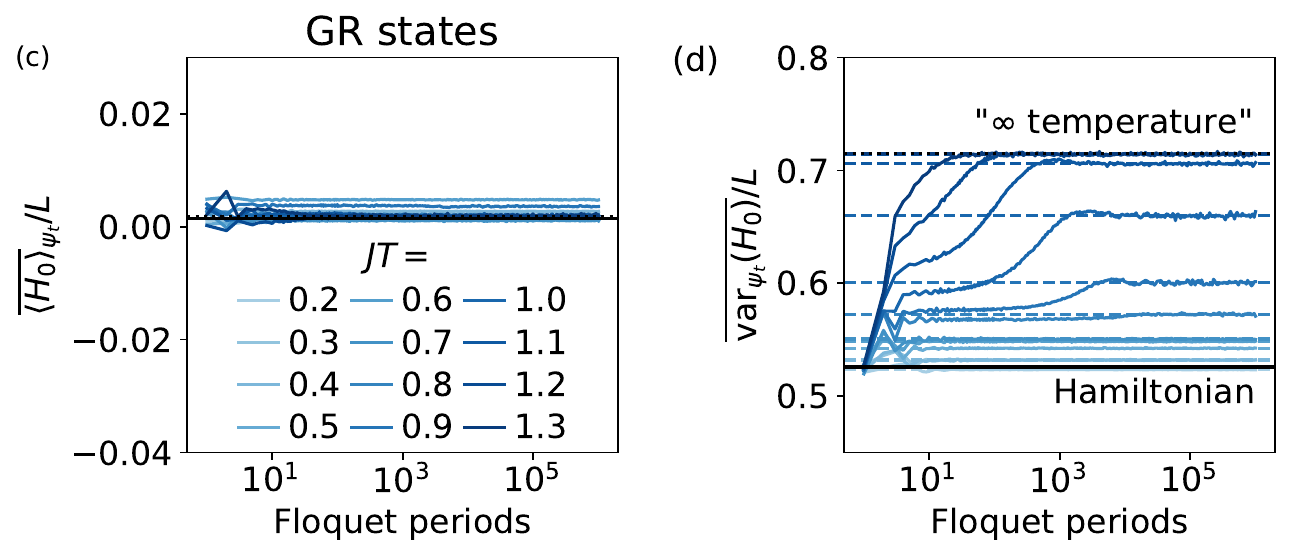}
{\color{blue}
\caption{Floquet heating dynamics. (a,b) IS and (c,d) GR states for different driving periods $JT$. The mean energy density (a,c) stays close to its infinite-temperature value throughout and is a poor heating diagnostic. The variance of $H_0$ (b,d) instead evolves away from its conserved Hamiltonian value (black solid line) toward the diagonal-ensemble prediction (colored dashed lines); the black dotted line marks the infinite-temperature value. Parameters: $J=1$, $h=0.9$, $N=16$, $\mathcal{N}=10^3$ initial states.}
\label{fig:heating dynamics}
}
\end{figure*}

Under $H_0$-only evolution, $\overline{\mathrm{var}_{\psi_t}(H_0)}$ is constant [Fig.\ \ref{fig:heating dynamics}(b,d)]. Under the exact Floquet map it instead evolves away from this value and saturates near the diagonal-ensemble prediction 
\begin{align}
  V_{\mathrm{DE}}&=\mathrm{Tr}(\rho_{\mathrm{DE}}H_0^2)-[\mathrm{Tr}(\rho_{\mathrm{DE}}H_0)]^2, 
  \\
   \rho_{\mathrm{DE}}&=\sum_n|\braket{E_n}{\psi}|^2\ket{E_n}\bra{E_n},
\end{align}
which accurately captures the long-time value. Both IS [Fig.\ \ref{fig:heating dynamics}(b)] and GR states [Fig.\ \ref{fig:heating dynamics}(d)] show qualitatively the same behavior. Complete
Floquet heating corresponds to the infinite-temperature
ensemble $\rho_\infty=\mathbf{1}/\mathcal{D}$ within the symmetry sector with Hilbert space dimension $\mathcal{D}$.  At finite $T$, the exact dynamics thus redistributes spectral weight relative to $H_0$ and cannot be identified with evolution under the static Hamiltonian.

\begin{figure*}[t!]
\centering
\includegraphics[width=0.3\linewidth]{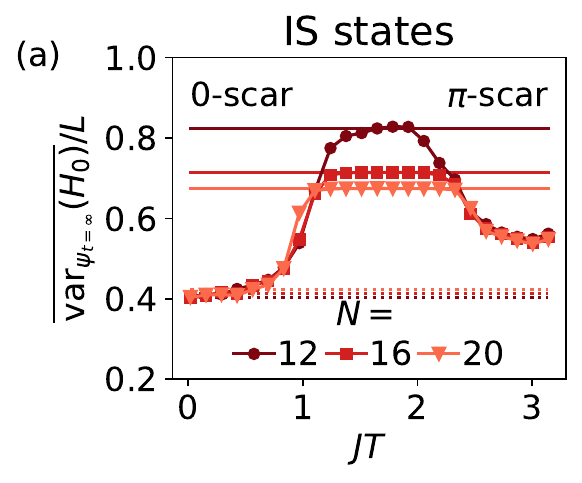}
\includegraphics[width=0.3\linewidth]{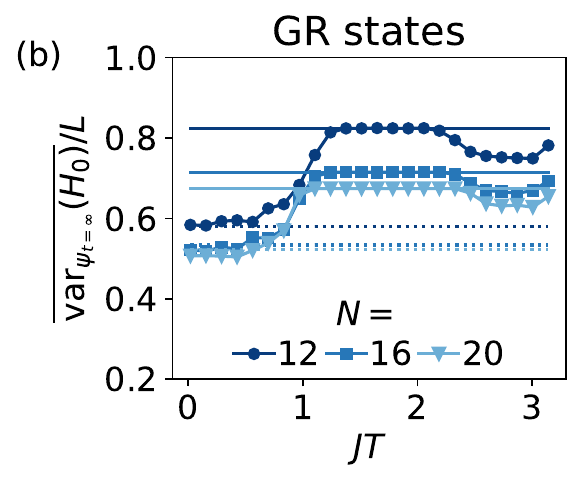}
\includegraphics[width=0.3\linewidth]{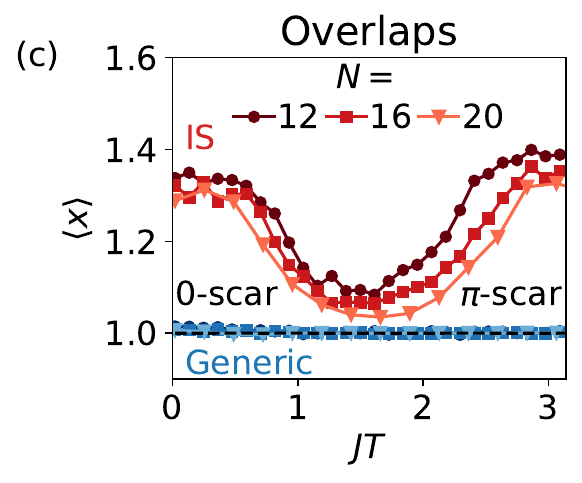}
{\color{blue}
\caption{Long-time Floquet heating and scarring vs.\ $T$. (a,b) Diagonal-ensemble variance of $H_0$ for IS and GR states, compared with the static-Hamiltonian and infinite-temperature values. Heating is incomplete for IS states in both the $0$- and $\pi$-scar regimes, and for GR states only in the $0$-scar regime. (c) Mean overlap $\langle x\rangle$: the scar excess is suppressed where heating is complete. The trend is the same for all system sizes studied. Parameters: $J=1$, $h=0.9$, $N=12,16,20$, $\mathcal{N}=10^3$.}
\label{fig:inf time heating}
}
\end{figure*}

Figure~\ref{fig:inf time heating}(a,b) shows the diagonal-ensemble variance vs.\ $T$, together with $\langle x\rangle$ (c). Starting from the Hamiltonian limit at small $T$, the stationary value departs from $H_0$ and approaches the infinite-temperature prediction near $JT\simeq1$. The heating crossover tracks the scar diagnostic: the scar excess remains finite wherever heating is incomplete, and both approach their infinite-temperature values in the same intermediate regime, for all system sizes studied. As noted above, in the $0$-scar corridor both IS and generic states resist heating, whereas in the $\pi$-scar corridor generic states are already thermal while IS states retain a finite-size deviation from $\rho_\infty$.
}

\section{System-size scaling}
\label{app:system-size scaling}
We perform a system-size scaling of the average overlap [Fig.\ \ref{fig:phase_diag} {\color{blue}(e)} and Fig.\ \ref{fig:end_matter_phase_diag} (d)]. While the enhancement of the overlaps in the scarred corridors persist for different $N$, they increasingly approach the RMT limit in the unscarred regime. The absence of curve crossings with system size indicates a crossover rather than a sharp transition. We interpret this crossover as a trotterization transition which appears when a Hamiltonian evolution is approximated by a sequence of discrete quantum gates with Trotter step $T$ \cite{heyl_quantum_2019,sieberer_digital_2019}. It is characterized by the localization properties of product states in the Floquet eigenbasis and can be diagnosed by the inverse participation ratio $\mathrm{IPR}=\sum_n x_n^2$. IS states have large overlaps on fewer eigenstates in the scarred corridors, but scramble across the full Hilbert space in the unscarred corridors.

{\color{blue}
\begin{figure}[t!]
    \centering
\includegraphics[width=0.6\linewidth]{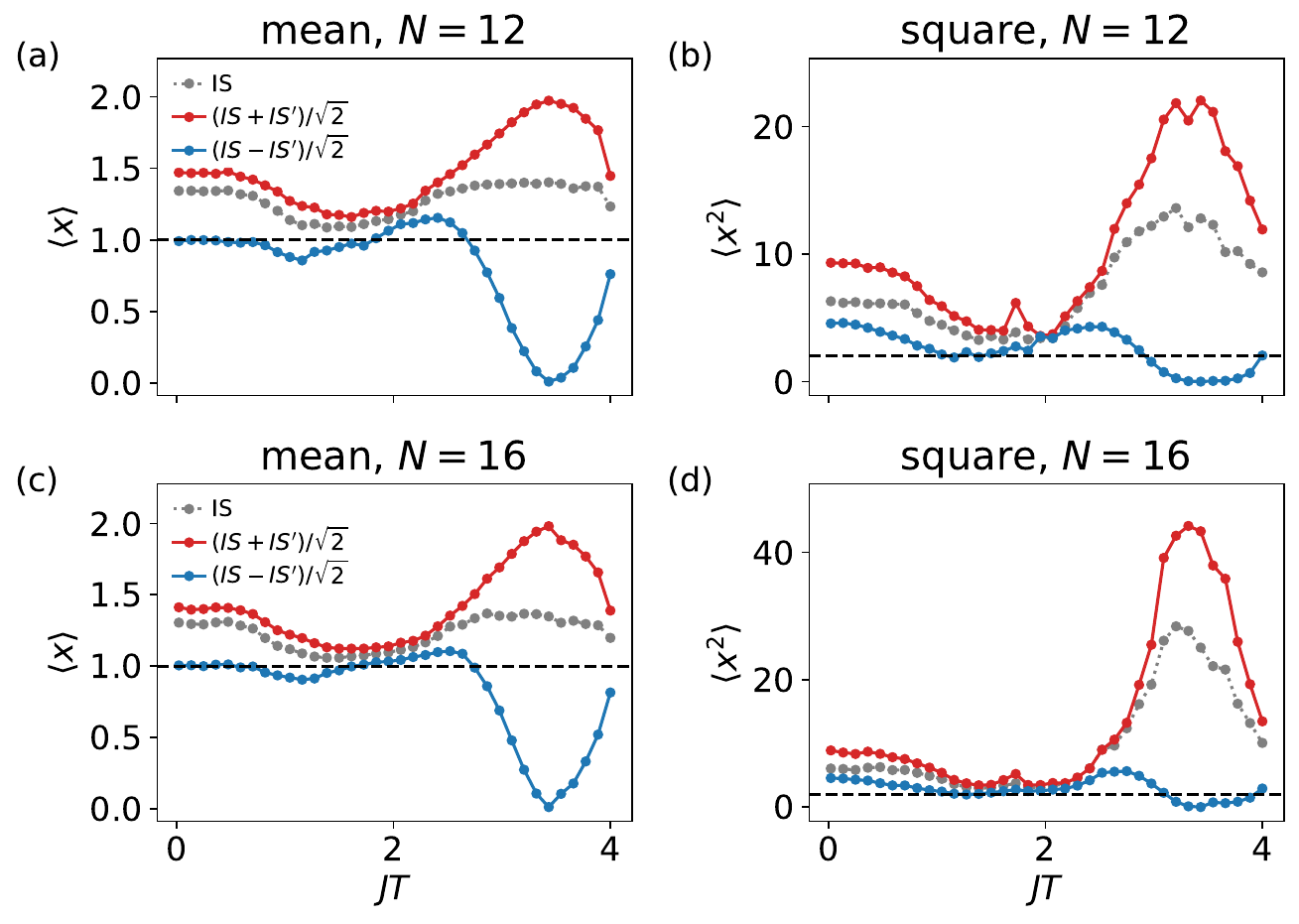}
    \caption{{\color{blue}
Overlap moments $\langle x\rangle$ [(a),(c)] and $\langle x^2\rangle$ [(b),(d)] vs.\ $T$ for IS, $\mathrm{IS}_+$, and $\mathrm{IS}_-$, at $N=12$ [(a),(b)] and $N=16$ [(c),(d)], using the central-$50\%$ quasienergy window of the main text. Parameters: $J=1$, $h=0.9$.
}
}
\label{fig:spin_flip_scaling}
\end{figure}

We repeat this scaling for the spin-flipped cat states at $N=12,16$ [Fig.\ \ref{fig:spin_flip_scaling}]. The enhancement of $\ket{\mathrm{IS}}_+$ over plain IS near the $\pi$-scar regime, and the corresponding suppression of $\mathrm{IS}_-$, persist at both system sizes, indicating that this cat-state signature of $\pi$-scars is not a finite-size artifact.
}

\end{document}